\documentclass[12pt]{iopart}

\usepackage{hyperref}
\begin{document}

\title{An entropy-driven cosmic expansion}
\author{F Hammad}
\address{D\'{e}partement ST2, Universit\'{e} A.Mira. Route Targa Ouzemmour, Bejaia 06000. Algeria}
\ead{fayhammad@gmail.com}

\begin{abstract}
We examine the evolution of the Friedmann Universe within our recent model of
space-time identified with an elastic continuous medium whose deformations
are described by a vector field constrained to obey a generalized
four-dimensional version of the equilibrium equations of standard
elasticity. It is found that the demand that the entropy associated with
such elastic deformations be always extremal during the expansion of such a
Universe turns these equilibrium equations into a single differential
equation governing the evolution of the Hubble parameter $H$. The solution
to the resulting dynamics admits both a power-law expansion, analogous to
the one induced by an inflaton field, as well as a power-law expansion
analogous to the one induced by a phantom field. Analyzing both types of
expansions via the induced elastic energy and pressure permits to assign
the former to the early Universe and the latter to its late-time expansion.
It is argued, however, that the present model does not exclude a phantom-like
inflation for the early Universe. We discuss the possible way for the dynamics
to avoid the Big Rip singularity that would otherwise result. We succinctly
discuss the possible way to avoid also the Big Bang singularity and how to obtain
the large scale structure of the Universe from the present model.
\end{abstract}



\maketitle

\section{Introduction}

\label{Intro} The Hot Big Bang theory, a theory that reproduces marvelously
all the actual data from the observed Universe, has led to the well-known
initial conditions problem (flatness problem, horizon problem, etc. See e.g.
\cite{1}). The idea of the inflationary scenario has rescued the Hot Big
Bang theory by providing a mechanism for producing such necessary initial
conditions \cite{2,3,4} as well as the observed large scale structure of the
present Universe \cite{5}. The idea is an early accelerated expansion of the
Universe, allowing the latter to achieve the needed initial conditions that
make it possible for its subsequent 'normal' expansion to present to us the
actually observed picture in the distribution of the CMBR (Cosmic Microwave
Background Radiation) in the sky. The idea of an early accelerated expansion
of the Universe is usually designated in literature by the word 'scenario'
because the details of its mechanism are model dependant since the inflaton
field responsible for generating such dynamics is a scalar field entering
into some specific Lagrangian models chosen in such a way as to reproduce
the precise needed rate and duration of the early expansion.

Another problem that rises within the Hot Big Bang theory when confronted
with the observed Universe is the presently observed accelerated expansion
of the latter \cite{6,7,8}. It is well-known that such expansion could
result at the present epoch if there were some kind of dark energy that
manifests itself through a cosmological constant leading to a late-time
accelerated expansion of the cosmos (see e.g. \cite{9}). But it is also
well-known that it is hard to come up naturally with a cosmological constant
that is positive and yet as small as the needed one to agree with
observations \cite{10}. Many models are also proposed in which one finds
another ingredient called a 'phantom' field \cite{11,12,13}. Indeed, the
possibility that the Universe might actually contain a kind of 'phantom'
dark energy is not ruled out by observation (see e.g. \cite{14,15}). The
phantom field may help reproduce the actually observed accelerated expansion
but may also lead to a Big Rip singularity in the Universe during a finite
time in the future.

Now, it would really be interesting if this idea of an accelerated expansion
of the Universe, both in its early times and in its later times,
appears as a manifestation of a fundamental principle of nature at work, and
intimately linked to the fabrics of space-time itself. The main goal of the
present paper is to motivate and examine such a possibility based on our
recent work on the idea of an elastic space-time continuum. In fact, many
authors have proposed and explored alternative ideas for generating such
inflationary and late-time expansions from a modified theory of space-time or
some other basic principle. Among many, we find for instance among the recent ones
the use of modified Hilbert-Einstein Lagrangian theories (see e.g. \cite
{16,17}), the recent proposal for explaining both inflation and the
late-time accelerated expansion in terms of an entropic force \cite{18,19},
and finally attributing the cosmic expansion to an elastic energy stored in
space-time and induced by the existence of a cosmic defect in the latter
\cite{20,21}. Our present approach based on the idea of an elastic
space-time and its associated entropy emerges from different motivations as
to how entropy might lead to a cosmic expansion (see also \cite{22}) and
stands on different constructions concerning how elasticity might enter the
dynamics.

Indeed, identifying space-time with an elastic continuous medium and then
constructing an entropy functional to be associated with its elastic
deformations has proved fertile in the sense that it permits to recover
straightforwardly both Einstein field equations and familiar results from
black hole thermodynamics \cite{23,24}. Einstein field equations emerged
from the second law of thermodynamics applied to the entropy functional,
i.e., they represent a constraint on any elastic space-time whose elastic
deformations always extremize the entropy associated with them. Black hole
thermodynamics formulas, i.e., the Hawking temperature and the
Bekenstein-Hawking entropy formula, are derived from the same entropy
functional when the deformation vector field $u^{i}$ contained in the latter
is constrained to obey a generalized four-dimensional version of the
equilibrium equations of standard elasticity \cite{24}. That is, the second
principle of thermodynamics is responsible for the emergence of the dynamics
as well as the thermodynamics of such space-times \footnote[1]{
The approach, when generalized to Riemann-Cartan space-times, also permits
to recover the Cartan-Sciama-Kibble field equations \cite{25}.}. We shall
see in the present paper that the same principle may also be behind the
dynamics of the observable Universe as a whole and may lead to the crucial
early accelerated expansion of the Universe without appealing to an inflaton
field and produce a late-time acceleration without appealing to a
cosmological constant or a phantom field.

The outline of this paper is as follows. In section~\ref{Friedmann}, we
recall the main ingredients that allowed us to arrive at Einstein field
equations by identifying space-time with an elastic continuum medium and
then use them to analyze the dynamics of the Friedmann Universe corresponding
to such a space-time. In section~\ref{Solutions}, we compute explicitly and
examine in detail the two solutions to the dynamical equation obtained for
the Hubble parameter $H$ in section~\ref{Friedmann}. In section~\ref
{Distinction}, we examine the elastic energy and pressure induced by the
elastic deformations and propose a scenario for how and why the Universe
might undergo an inflationary expansion early in its history before reaching
a radiation and then matter dominated eras. We then discuss how it might
have entered its late-time evolution that looks like a phantom dark
energy-dominated era. We end this paper with a general conclusion discussing
the possible way to avoid the Big Bang singularity and the way the large
scale structure of the Universe could emerge from the model.

\section{Entropy and the dynamics of a Friedmann elastic Universe}\label{Friedmann}
In our elastic space-time approach \cite{24}, following
ref.~\cite{23}, we considered space-time to be a continuum medium whose
elastic deformations are quantified by the vector field $u^{i}=\overline{x}
^{i}-x^{i}$, where $i=0,...,3$ and $\overline{x}^{i}$ and $x^{i}$ are
coordinate labels in the space-time after and before deformation,
respectively. An entropy, to be associated with these deformations, is then constructed as a scalar quadratic in the first derivatives of the field $\partial u$ -- in analogy with the usual thermodynamic potentials found in standard elasticity (see e.g. \cite{26}) -- as well as in the field $u^{i}$ itself. The latter contribution, usually not found in standard elasticity, comes from the manifestation of the breaking of translational invariance due to matter viewed in this approach as dislocations in the space-time medium \cite{23}. The
most general covariant form obtained for the functional was \cite{23,24} {\footnote[2]{We shall adopt through out this paper the natural units $\hbar=c=1$}
\begin{equation}
\mathcal{S}=\frac{1}{8\pi G}\int_{\mathcal{M}}\mathrm{d}^{4}x\sqrt{-g}\Big[
\nabla _{i}u_{j}\nabla ^{j}u^{i}-(\nabla _{i}u^{i})^{2}+8\pi G(\frac{1}{2}
g_{ij}T-T_{ij})u^{i}u^{j}\Big].  \label{1}
\end{equation}
Here, the eventual cosmological constant that the functional (\ref{1}) might
contain is omitted in order to investigate the dynamics of the Universe
without a cosmological constant. This functional was then varied with
respect to the field $u^{i}$ (held fixed at the boundaries) by imposing the extremality condition $\delta \mathcal{S}=0$ for all possible deformations $u^{i}$ in the bulk. The resulting equations are nothing but the
Einstein field equations $R_{ij}-\frac{1}{2}g_{ij}R=-8\pi GT_{ij}$. Since during the variation performed on the functional (1) the field $u^{i}$ was not allowed to vary on the boundaries, when the emerging Einstein equations are substituted back into (1) and the latter is integrated by parts the bulk degrees of freedom cancel away leaving only the degrees of freedom on the boundaries left out during the variation. Indeed, substituting these back into the original functional (\ref{1}) and then integrating by parts gives an 'on-shell' entropy functional that reads \cite {23}
\begin{equation}
\mathcal{S}=\frac{1}{8\pi G}\int_{\partial \mathcal{M}}\mathrm{d}^{3}x\sqrt{
|h|}n_{i}(u^{j}\nabla _{j}u^{i}-u^{i}\nabla _{j}u^{j}),  \label{2}
\end{equation}
where $h$ is the determinant of the three-metric on the hypersurface
bounding the integrated region of space-time. At this point, nothing is
imposed yet on the vector field $u^{i}$; Einstein equations emerge only from
imposing $\delta \mathcal{S}=0$ for whatever values the field $u^{i}$ might
happen to take. That is, the construction suggests that the dynamics of the
metric of space-time (i.e., gravity) emerges from a deeper level in the
structure of space-time at which the latter always tends to satisfy the
second law of thermodynamics. However, taking this vector field to be a kind
of 'vector-state' representing the difference between the two states of
space-time after and before deformation as in elasticity theory, naturally
suggests to picture each infinitesimal volume element of space-time as if it
were in equilibrium within the whole continuum due to the elastic stresses
it experiences from its surrounding neighbors. The simplest (and heuristic)
way to express the equilibrium condition of these elements of space-time is
then to adopt a four-dimensional generalization of the Hooke's law
$\sigma_{\alpha \beta }=\mu \delta _{\alpha \beta }{\varepsilon _{\gamma }}^{\gamma}+2\nu \varepsilon _{\alpha \beta }$,
giving the relation between the stress tensor $\sigma _{\alpha \beta }=\sigma _{\beta \alpha }$
and the strain tensor $\epsilon _{\alpha \beta }=\frac{1}{2}(\partial _{\alpha }u_{\beta}+\partial _{\beta }u_{\alpha })$,
as well as a four-dimensional generalization of the usual equilibrium equations of elasticity
$\partial_{\alpha }\sigma ^{\alpha \beta }=0$ expressed in terms of the stress tensor
in the absence of external and non-elastic forces ($\alpha ,\beta =1,2,3$)
(see e.g. \cite{26}). In standard elasticity, the above form of the Hooke's
law applies whenever the continuum in consideration is homogeneous and
isotropic as we shall suppose it to be the case for space-time \footnote[3]{
For an elaborate discussion on the tensors of anisotropic elasticity see
\cite{27}.}. The
positive coefficients $\mu $ and $\nu $ are called the Lam\'{e} coefficients
and describe the elastic properties of the continuum. The stress tensor
$\sigma ^{\alpha \beta }$ represents the force (per unit area) applied
on an infinitesimal volume element of the continuum, pointing in the direction
$\alpha $ and perpendicular to a surface whose normal is in the direction
$\beta $. In addition, when space-time is not considered as embedded in a
higher-dimensional manifold, one may also want to generalize the vanishing
of the rigid-rotations tensor $\omega _{\alpha \beta }=\frac{1}{2}(\partial
_{\alpha }u_{\beta }-\partial _{\beta }u_{\alpha })$ in ordinary elasticity
to four dimensions. The equations one obtains when all these ingredients are
transcribed into a covariant four-dimensional form are \cite{24}
\begin{equation}
\sigma ^{ij}=\mu g^{ij}\nabla _{k}u^{k}+2\nu \nabla ^{i}u^{j},  \label{3}
\end{equation}
for the generalized Hooke's law, and
\begin{equation}
\mu \big(\nabla ^{i}\nabla _{j}u^{j}\big)+2\nu \big(\nabla _{j}\nabla
^{i}u^{j}\big)=0,  \label{4}
\end{equation}
for the generalized equilibrium equations. Here $\mu $ and $\nu $ would be
the positive Lam\'{e} elastic coefficients of space-time. Using the Ricci
identity $[\nabla _{j},\nabla _{i}]u^{j}=R_{ij}u^{j}$ in curved space-times,
where the symmetric Ricci tensor $R_{ij}$ is constructed from the metric
$g_{ij}$ and the Riemann tensor $R_{ij}=g^{kl}R_{kilj}$, the above equations
also read
\begin{equation}
\nabla _{i}\nabla _{j}u^{j}=-\frac{2\nu }{\mu +2\nu }R_{ij}u^{j}.  \label{5}
\end{equation}
In standard elasticity, the Lam\'{e} coefficients depend on temperature and
the density of defects inside the elastic material. In our approach, matter
and pure energy are viewed as dislocations (or defects) in the space-time
continuum. As such, we may also expect non-trivial modifications of these
equilibrium equations when the density of defects (i.e., matter or radiation)
becomes non-negligible. We shall discuss below in detail the consequences of
this assumption within the framework of the present model.

All that remains now is to specify the space-time we wish to study and apply
these two constraints, namely, that its entropy due to its elastic
deformations must be given by (\ref{2}) in order for it to have a constantly
extremized value and its deformations obey (\ref{5}) in order for it to be a
continuum in equilibrium at everyone of its points. The specific space-time
we shall consider hereafter is the homogenous and isotropic Friedmann
Universe described by the Friedmann-Lema\^{\i}tre-Robertson-Walker (FLRW)
metric
\begin{equation}
\mathrm{d}l^{2}=-\mathrm{d}t^{2}+a^{2}(t)\Big(\frac{\mathrm{d}r^{2}}{1-kr^{2}
}+r^{2}\mathrm{d}\theta ^{2}+r^{2}\sin ^{2}\theta \mathrm{d}\varphi ^{2}\Big).  \label{6}
\end{equation}
$a(t)$ is the time-dependent scale factor and $k=-1,0,+1$ is the parameter
that distinguishes the open, flat, and closed Universes, respectively. Now
in order to deduce the dynamics that governs the time evolution of such a
space-time, i.e., the scale factor $a(t)$, one usually injects into the
left-hand side of Einstein equations the Einstein tensor $R_{ij}-\frac{1}{2}
g_{ij}R$ that comes out from the selected space-time (\ref{6}) while in the
right-hand side one injects the energy-momentum tensor $T_{ij}=\mathrm{diag}
(\rho ,-p,-p,-p)$ that comes from the radiation or the matter filling that
space-time. In the resulting dynamics, described by the Friedmann-Lema\^{\i}tre
equations $\dot{a}^{2}/a^{2}=\frac{8\pi G}{3}\rho -\frac{k}{a^{2}}$ and
$\ddot{a}/a=-\frac{4\pi G}{3}(\rho +3p)$, the density $\rho $ and the
pressure $p$ of matter or radiation are the unique sources capable of
inducing an expansion for the Universe. The radiation-dominated era (for
which the equation of state is $p/\rho =w=\frac{1}{3}$) gives $a(t)\sim
t^{1/2}$ whereas the matter (in the form of dust)-dominated era (for which
$p/\rho =w=0$) yields $a(t)\sim t^{2/3}$. The two equations may actually be
combined after computing the first derivative of each and then using the
definition $\dot{a}/a=H$ of the Hubble parameter as well as the identity
$\ddot{a}/a=\dot{H}+H^{2}$. The result is the following single differential
equation to which we shall refer below,
\begin{equation}
-\frac{\ddot{H}}{H^{3}}-(5+3w)\frac{\dot{H}}{H^{2}}=3(1+w).  \label{7}
\end{equation}
Thus, for all three possible values of $k$, the Hubble parameter $H$ cannot
induce a power-law expansion for $a(t)$ of the form $A/t$ in early-times or
$B/(t_{0}-t)$ (with $t_{0}>t$) in later-times unless the equation of state of
the sources entering (\ref{7}) is $-1<w<0$ or $w<-1$, respectively. The
first solution gives rise to what is known as a power-law inflation, whereas
the second induces the phantom behavior. This is where the need for an
inflaton and then a phantom field comes from. Given that our approach
consisting of assigning to space-time an entropy related to its elastic
deformations and which, when extremized, yields its dynamics it is tempting
to think that this same approach may help follow the dynamics of space-time
through its whole evolution beginning from its early times immediately after
the Planck era and all the way through to its late-time expansion without
appealing to inflaton fields for early times and phantom fields for later
times. In other words, instead of investigating the evolution of the
Universe using Einstein equations (i.e., the 'metric' gravity) that tell us
how matter and energy when they dominate influence the dynamics of the
metric through the Friedmann-Lema\^{\i}tre equations we, in a sense, descend
into a deeper level and check directly the deformations of space-time (the
'entropic' elastic gravity) that gave birth in the first place, through the
second law of thermodynamics, to the dynamics of the metric. At this level
it is legitimate, in principle, to expect that there would be no need for
particular treatments for each of the four different epochs in the history
of the Universe. We shall come back to this last point later.

We begin then from the usual assumptions of homogeneity and isotropy of
space. First, from the assumption of homogeneity of space we expect only
time-dependent components of the field $u^{i}$ whereas isotropy suggests to
have only a time-component $u^{0}$ of the field $u^{i}$ and possibly also a
radial component $u^{1}$. That is, we shall start with the following ansatz
for the deformation vector field
\begin{equation}  \label{8}
u^{i}=(u^{0},u^{1},0,0)\equiv(\phi(t),\varrho(t),0,0).
\end{equation}
As a consequence, the resulting 'on-shell' entropy takes the following form
\begin{eqnarray}  \label{9}
\mathcal{S}=-\frac{1}{8\pi G}\int\frac{a^{3} r^{2}\sin\theta}{\sqrt{1-kr^{2}}}\Big(u^{1}\nabla_{1}u^{0}-u^{0}\nabla_{1}u^{1}\Big)\mathrm{d}r\mathrm{d}\theta\mathrm{d}\varphi  \nonumber \\
\qquad+\frac{1}{8\pi G}\int\frac{a^{3} r^{2}\sin\theta}{\sqrt{1-kr^{2}}}\Big(u^{0}\nabla_{0}u^{1}-u^{1}\nabla_{0}u^{0}\Big)\mathrm{d}t\mathrm{d}\theta\mathrm{d}\varphi.
\end{eqnarray}

Next, let us start from the FLRW metric (\ref{6}) and obtain the precise
equilibrium equations that result for the field $u^{i}$. Although there are
in (\ref{5}) four equilibrium equations in all, due to the form (\ref{8}) of
the deformation field only two equations remain. They are
\begin{equation}  \label{10}
\partial_{t}(\dot{\phi}+\Gamma_{j0}^{j}\phi+\Gamma_{j1}^{j}\varrho)=-\frac{2\nu}{\mu+2\nu}(R_{00}\phi+R_{01}\varrho),
\end{equation}
\begin{equation}  \label{11}
\partial_{r}(\dot{\phi}+\Gamma_{j0}^{j}\phi+\Gamma_{j1}^{j}\varrho)=-\frac{2\nu}{\mu+2\nu}(R_{10}\phi+R_{11}\varrho).
\end{equation}
The needed non-vanishing Christoffel symbols $\Gamma_{\alpha0}^{\alpha}$ and
$\Gamma_{\alpha1}^{\alpha}$ ($\alpha=1,2,3$) as well as the non-vanishing
components $R_{00}$ and $R_{11}$ of the Ricci tensor are extracted from the
FLRW metric (\ref{6}). They are, respectively,
\begin{equation}  \label{12}
\Gamma_{10}^{1}=\Gamma_{20}^{2}=\Gamma_{30}^{3}=\frac{\dot{a}}{a}\quad,\quad\Gamma_{21}^{2}=\Gamma_{31}^{3}=\frac{1}{r},
\end{equation}
\begin{equation}  \label{13}
R_{00}=\frac{3\ddot{a}}{a}\quad,\quad R_{11}=-\frac{a\ddot{a}+2\dot{a}^{2}+2k}{1-kr^{2}}.
\end{equation}
Substituting these in (\ref{10}) and (\ref{11}), the equilibrium equations
take the following more explicit expressions
\begin{equation}  \label{14}
\partial_{t}(\dot{\phi}+3\frac{\dot{a}}{a}\phi+\frac{2}{r}\varrho)=-\frac{6\nu}{\mu+2\nu}\frac{\ddot{a}}{a}\phi,
\end{equation}
\begin{equation}  \label{15}
\partial_{r}(\dot{\phi}+3\frac{\dot{a}}{a}\phi+\frac{2}{r}\varrho)=\frac{2\nu}{\mu+2\nu}\frac{a\ddot{a}+2\dot{a}^{2}+2k}{1-kr^{2}}\varrho.
\end{equation}

Now, performing the $r$-derivative in the left-hand side of the last
equation and then transposing the denominator $1-kr^{2}$ in the right-hand
side to the left, we immediately see that the only way for the two sides --
one having a multiplicative factor that depends only on the variable $r$
whereas that of the other depends only on the variable $t$ -- to be equal is
to have an identically vanishing $\varrho (t)$. So we conclude that the
field $u^{i}$ that is compatible with the FLRW geometry and simultaneously
constrained to obey the generalized equilibrium equations (\ref{4}) should
be of the form $u^{i}=(\phi (t),0,0,0)$. The only remaining constraint to be
imposed on this field is therefore Eq.~(\ref{14}). Performing the
time-derivative and using $H=\dot{a}/a$ and $\ddot{a}/a=\dot{H}+H^{2}$ in
order to recast the equation in terms of the Hubble parameter $H$ instead of
the scale factor $a$, the constraint equation becomes the following
dynamical equation for $\phi $
\begin{equation}
\ddot{\phi}+3H\dot{\phi}+\Big(\frac{3\mu +12\nu }{\mu +2\nu }\dot{H}+\frac{6\nu }{\mu +2\nu }H^{2}\Big)\phi =0.  \label{16}
\end{equation}
We notice a form similar to the familiar dynamical equations usually
obtained for the scalar $\phi $ in the scalar field models approach to
inflation and phantom dark energy, but with a potential energy that is a $H$
and $\dot{H}$-dependent function. Going back to (\ref{9}), the 'on-shell'
functional associated with the field $(\phi (t),0,0,0)$ simplifies further
to
\begin{equation}
\mathcal{S}=\frac{3\dot{a}a^{2}{\phi }^{2}}{8\pi G}\int \frac{r^{2}\sin\theta }{\sqrt{1-kr^{2}}}\mathrm{d}r\mathrm{d}\theta \mathrm{d}\phi =\frac{3
\mathcal{V}_{r}}{8\pi G}\,\dot{a}a^{2}{\phi }^{2},  \label{17}
\end{equation}
where $\mathcal{V}_{r}$ is the constant $\int\frac{r^{2}\sin \theta }{\sqrt{1-kr^{2}}}\mathrm{d}r\mathrm{d}\theta\mathrm{d}\phi$
representing the co-moving coordinate-volume. We see from this functional that in order for
the entropy associated with the elastic deformations of space-time to be
strictly positive and always extremal, in accordance with the second law of
thermodynamics, the scale factor should be non-vanishing and increasing with
time to insure $\dot{a}>0$. Furthermore, the scalar field $\phi $ should
absorb any such increase of the scale factor $a$ as well as its derivative $
\dot{a}$ in (\ref{17}). That is, the scalar $\phi $ should take the
following form
\begin{equation}
\phi \sim \frac{1}{a\sqrt{\dot{a}}}=\frac{1}{a^{3/2}\sqrt{H}},  \label{18}
\end{equation}
the Hubble parameter $H$ being strictly positive as it follows from the
argument above. Substituting this form of the field $\phi $ in Eq.~(\ref{16}),
the latter becomes the following nonlinear second order differential equation for the Hubble parameter $H$
\begin{equation}
-\frac{\ddot{H}}{H^{3}}+\frac{3}{2}\frac{\dot{H}^{2}}{H^{4}}+3\lambda \frac{\dot{H}}{H^{2}}=\frac{3}{2}\eta ,  \label{19}
\end{equation}
where $\lambda=(\mu+6\nu)/(\mu+2\nu)$ and $\eta=(3\mu-2\nu)/(\mu+2\nu)$.
This is the equation that governs, in our model, the dynamics of
the elastic Friedmann Universe. It looks like Eq.~(\ref{7}) obtained from the
Friedmann-Lema\^{\i}tre equations, except for the one additional term $\dot{H}^{2}/H^{4}$.
However, in contrast to Eq.~(\ref{7}) this last equation allows
solutions for the Hubble parameter $H$ of the form $A/t$ as well as $
B/(t_{0}-t)$ without introducing matter sources whose equations of state
should satisfy $w<0$. The trick is done by the additional term $\dot{H}
^{2}/H^{4}$ as well as the Lam\'{e} coefficients whose role is to influence
the behavior of the elastic deformations.

Another peculiar feature of the dynamics induced by Eq.~(\ref{19}) is its explicit independence on the parameter $k$ that distinguishes the closed, flat, and open Universes. Referring to Eq.~(\ref{7}), this fact can be understood as follows. Although Eq.~(\ref{7}) does not display explicitly the parameter $k$, either, it does actually depend on the latter through the parameter $w$. In fact, $w$ comes from the source's equation of state and is related to the parameter $k$ through the Friedmann-Lema\^{\i}tre equations. Solving for $H$ after fixing $w$ in (\ref{7}) by choosing the source -- matter, radiation, scalar fields, etc. -- with a specific equation of state automatically fixes the geometry through the Friedmann-Lema\^{\i}tre equations. The dynamics induced by Eq.~(\ref{19}), however, does not result from an external source but comes exclusively from space-time itself. The latter would indeed have been sensitive to the parameter $k$ through Eq.~(\ref{15}) if the scalar $\phi$ were not a homogeneous field. On the other hand, the fact that the contribution of the parameter $k$ in the final entropy formula (\ref{17}) factors out in the form of a constant implies that the dynamics of space-time, which is mainly driven by the tendency of entropy to remain extremal, does not in the end distinguish between the closed, flat, and open geometries.

In the next section, we examine the detailed form of the two solutions $H=A/t$ and $H=B/(t_{0}-t)$, relate them respectively to the inflationary expansion and the late-time expansion of the Universe, and then discuss the intermediate radiation and
matter-dominated eras.

\section{From inflation to late-time acceleration}

\label{Solutions}

\subsection{Inflationary expansion}

When $\eta \neq 0$, Eq.~(\ref{19}) admits the following two solutions
$H=A/(t_{0}+t)$ and $H=B/(t_{0}-t)$ for some arbitrary constants $A$, $B$, and
$t_{0}$. The case $\eta =0$ will be discussed in the last section of this
paper. We begin here by examining the first one after shifting the time
$t_{0}+t\rightarrow t$ for simplicity. Inserting $H=A/t$ into (\ref{19}) we
find two possible values for $A$:
\begin{equation}
A=\frac{-3\lambda \pm \sqrt{9\lambda ^{2}-3\eta }}{3\eta }.  \label{20}
\end{equation}
Remembering, though, that $\eta =(3\mu -2\nu )/(\mu +2\nu )$, we have also
to distinguish between $\eta <0$ when $\mu <2\nu /3$ and $\eta >0$ when $\mu
>2\nu /3$. For positive $\eta $, however, the two solutions (\ref{20}) both
yield negative values for the constant $A$ and hence also negative values
for the Hubble parameter $H$. Thus, only $\mu <2\nu /3$ permits to have $H>0$
and one must choose the solution with the minus sign in (\ref{20}).
Furthermore, when $|\eta |\ll 1<\lambda $ the following approximation
$A\approx -2\lambda /\eta $ holds, and the corresponding time-dependence of
the scale factor is
\begin{equation}
a(t)\propto t^{-\frac{2\lambda}{\eta}}=t^{\frac{2(\mu+6\nu)}{2\nu-3\mu}}.  \label{21}
\end{equation}
We see that we can get a sufficient inflationary expansion within this
model, and more precisely the required minimum amount of 70 e-foldings \cite{5},
provided that the Lam\'{e} coefficients $\mu$ and $\nu$ satisfy $32\nu/53\leq\mu<2\nu/3$. Now although the latter condition exhibits a fine-tuning character its meaning is actually less disturbing when viewed as a condition on the Poisson's ratio $\varsigma=\mu/[2(\mu+\nu)]$. Indeed, recall that in standard elasticity \cite{26} the latter measures the ratio of the transverse compression to the longitudinal extension of the medium, and it is constrained to take values within the interval $0<\varsigma<0.5$ when the Lam\'{e} coefficient $\mu$ is positive. It is well-known in solid state physics that this ratio differs from one material to the other, being dependent on the atomic structure of the medium. In our case the above condition becomes simply $0.19\leq\varsigma<0.2$, and hence, rather than representing a fine-tuning, it is nothing but an indication on a physical characteristic of the continuous medium, namely one of its elastic properties. It is easy to see that the interval $0<\varsigma<0.2$ permits actually to recover from (21) all the preferred exponents $-\frac{2\lambda }{\eta }\gg10$ of power-law inflation \cite{28}. Next, we shall investigate the implication
of this approximation on the other solution, namely when $H=B/(t_{0}-t)$.

\subsection{Late-time expansion}

Inserting the other solution $H=B/(t_{0}-t)$ into Eq.~(\ref{19}) we also
find two possible values for $B$:
\begin{equation}
B=\frac{3\lambda \pm \sqrt{9\lambda ^{2}-3\eta }}{3\eta }.  \label{22}
\end{equation}
This time, however, both $\eta >0$ and $\eta <0$ permit to have positive
values for $B$ and hence a positive $H$. Working with the fixed choice $\mu<2\nu/3$
that gives a negative $\eta $, and hence a positive $H=A/t$ for
the first solution, we must choose the minus sign in (\ref{22}) in order to
have again a positive $H$ for this case. We verify then that the
approximation $|\eta |\ll 1<\lambda $ that allows us to find an inflationary
expansion gives the moderate coefficient $B\approx1/(6\lambda)$, which in
turn induces the following time-dependence for the scale factor
\begin{equation}
a(t)\propto \frac{1}{(t_{0}-t)^{\frac{1}{6\lambda }}}=\frac{1}{(t_{0}-t)^{\frac{\mu +2\nu }{6(\mu +6\nu )}}}.  \label{23}
\end{equation}
For the case $32\nu/53\leq\mu<2\nu/3$ this becomes, in the lower limit,
approximately $a\propto (t_{0}-t)^{-0,07}$. The exponent is, however, not
sufficient compared to what actual observations suggest for the phantom
power-law \cite{29}, but this is the best we can achieve in this model with
fixed Lam\'{e} coefficients. Expression (\ref{23}) obviously implies the Big
Rip singularity in the finite time $t_{0}$ for any positive value of $\lambda$.
We shall come back to this fact below. But for now it is
satisfactory to have found both the inflaton field-induced behavior as well
as the phantom dark energy one within the same dynamical equation and within
the same approximation scheme for both.

\subsection{Radiation and matter-dominated eras}

As for getting the intermediate radiation and matter-dominated eras, one might
also be tempted to seek a solution for the corresponding dynamics using the
same equation (\ref{19}). However, identifying the latter with Eq.~(\ref{7})
which is valid for these eras leads one to impose negative values on the Lam\'{e}
coefficients and, even more, different ones for each of the two eras.
The reason for the failure of Eq.~(\ref{19}) to take into account the
presence of matter and radiation is the fact that, as we have alluded to it
in section~\ref{Friedmann}, matter and radiation are viewed as defects in the
space-time continuum that spoil the usual equilibrium equations of
elasticity in their neighborhoods. To find the deformation field in the
presence of these requires taking into account the presence of defects when
writing down the equilibrium equations themselves and this should be done,
not within a generalization of the linear three-dimensional elasticity
theory, but within the framework of a generalized version of the theory of
defects in crystals \cite{30}. That can also, in principle, be done by
building a model in which the density of defects of space-time intervenes in
the equilibrium equations. That, however, is beyond the scope of our present
model, and so the latter still requires the Friedmann-Lema\^{\i}tre equations
(i.e., the 'metric' gravity) to treat radiation and matter-dominated eras
properly.

\section{Distinguishing inflation from late-time expansion}\label{Distinction}
At this point, and using Eq.~(\ref{19}) alone, there is
yet no way to distinguish between the dynamics of the early-times and the
dynamics of the late-times. So there is no reason to assign the solution
$H=A/t$ to the former and $H=B/(t_{0}-t)$ to the latter. Indeed, one may very
well take instead the first solution to represent a late-time quintessential
power-law expansion \cite{17,29}, but then one is left with no other choice
for early-time inflation except to adopt the phantom-like behavior (\ref{23}).
Examining the elastic energy of space-time as well
as the elastic pressure, both induced by the elastic deformations, provides
complementary information on the dynamics, however.

When generalizing the symmetric stress tensor $\sigma^{\alpha\beta}$
($\alpha, \beta=1,2,3$) of standard elasticity to four dimensions, the
resulting symmetric tensor $\sigma^{ij}$ acquires the new components
$\sigma^{\alpha 0}$ and $\sigma^{00}$ that we don't find in three dimensions.
Since $\sigma^{\alpha\beta}$ represents in standard elasticity a force per
unit area, applied in the direction $\alpha$ and perpendicular to the
surface element whose normal is in the direction $\beta$, we shall interpret
the components of the generalized stress tensor as follows. The component
$\sigma^{\alpha\beta}$ would represent the force per unit area, applied in
the direction $\alpha$ and perpendicular to the elementary hypersurface
$\mathrm{d}x^{2}\mathrm{d}t$ whose normal is in the direction $\beta$. That
is, $\sigma^{\alpha\beta}$ is the pressure per unit cosmic time on a two
dimensional surface. The component $\sigma^{\alpha 0}$ would represent the
energy contained in an elementary hypersurface $\mathrm{d}x^{2}\mathrm{d}t$
whose normal is in the direction $\alpha$. That is, $\sigma^{\alpha 0}$ is a
flux of energy through a two dimensional surface. The component $\sigma^{00}$
then would be the energy stored inside the elementary hypersurface
$\mathrm{d}x^{3}$ whose normal is in the time direction, i.e., the three-dimensional
spatial volume $\mathrm{d}V$. That is, $\sigma^{00}$ is the elastic energy
density. Using Eq.~(\ref{3}), the non-vanishing Christoffel symbols (\ref{12}),
and the expression (\ref{18}) for the field $\phi$, we arrive at the
following identities that are also expressed in terms of the scale factor $a$
and the Hubble parameter $H$
\begin{equation}  \label{24}
\sigma^{\alpha 0}=0,
\end{equation}
\begin{equation}  \label{25}
p\equiv\frac{1}{3}{\sigma_{\alpha}}^{\alpha}=\mu\dot{\phi}+(3\mu+2\nu)H\phi=\frac{\sqrt{H}}{2a^{3/2}}\Big[\mu(3-\frac{\dot{H}}{H^{2}})+4\nu\Big],
\end{equation}
\begin{equation}  \label{26}
\mathcal{E}\equiv\sigma^{00}=-(\mu+2\nu)\dot{\phi}-3\mu H\phi=\frac{-\sqrt{H}}{2a^{3/2}}\Big[\mu(3-\frac{\dot{H}}{H^{2}})-2\nu(3+\frac{\dot{H}}{H^{2}})\Big].
\end{equation}

We first notice the vanishing of the flux of energy as perceived in a
co-moving reference frame through any two dimensional spatial surface, as
one would expect from our assumption that space-time is not imbedded in a
higher dimensional manifold. Indeed, one might expect possible energy flows only
if there were extra-dimensions outside the four-dimensional space-time; an
eventuality that we won't consider here. The energy density and pressure
are, however, not null and we will compute their corresponding values
separately for each of the two solutions found for $H$ from Eq.~(\ref{19}).

For $H=A/t$ we have $\dot{H}=-H^{2}/A$ with $A$ given by (\ref{20}) which,
in the approximation $|\eta|\ll\lambda$, reduces to $-2\lambda/\eta$, so
that $\dot{H}\approx\eta H^{2}/2\lambda$. Substituting this in the above
expressions for pressure and energy density, these read
\begin{equation}  \label{27}
p\approx\frac{\sqrt{H}}{2a^{3/2}}(3\mu+4\nu),
\end{equation}
\begin{equation}  \label{28}
\mathcal{E}\approx\frac{-\sqrt{H}}{2a^{3/2}}(3\mu-6\nu).
\end{equation}
Recalling that $H$ comes out positive when $\mu<2\nu/3$, we notice that both
expressions above are positive. Hence, for the solution $H=A/t$ the pressure
pushes outwards and the energy density is positive.

For $H=B/(t_{0}-t)$ we have $\dot{H}=H^{2}/B$ with $B$ given by (\ref{22})
which, in the approximation $|\eta |\ll \lambda $, reduces to $1/(6\lambda)$
, so that $\dot{H}\approx 6\lambda H^{2}$. Substituting this in the above
expressions for pressure and energy density, these read
\begin{equation}
p=-\frac{\sqrt{H}}{2a^{3/2}}\:\frac{3\mu ^{2}+26\mu \nu -8\nu ^{2}}{\mu +2\nu},  \label{29}
\end{equation}
\begin{equation}
\mathcal{E}=\frac{\sqrt{H}}{2a^{3/2}}\:\frac{3\mu ^{2}+48\mu \nu +84\nu ^{2}}{\mu +2\nu }.  \label{30}
\end{equation}
We notice that the energy density is positive whereas the pressure comes out
negative in the range $32\nu/53\leq\mu <2\nu/3$. Hence, for the solution
$H=B/(t_{0}-t)$ the pressure pulls inwards even if the energy density is
positive and the Universe expands. Before drawing the corresponding picture
for the two accelerated phases of the cosmic expansion that emerges from
this analysis, we also compute the total elastic energy $E=\int \mathcal{E}
\mathrm{d}V=a^{3}\mathcal{V}_{r}\mathcal{E}$ from the above energy densities
corresponding to each solution. Omitting the constant factors, we write only
the time-dependence of these and display beside each one of them the
corresponding time-dependence of the energy-density and the field $\phi\sim a^{-3/2}H^{-1/2}$.
For $H=A/t\approx -2\lambda /(\eta t)$, we get
\begin{equation}
E\sim t^{\frac{3\lambda }{|\eta |}},\qquad \mathcal{E}\sim t^{-\frac{
3\lambda }{|\eta |}},\qquad \phi \sim t^{-\frac{3\lambda }{|\eta |}}.
\label{31}
\end{equation}
While for $H=B/(t_{0}-t)\approx 1/[6\lambda (t_{0}-t)]$, we find
\begin{equation}
E\sim (t_{0}-t)^{-\frac{1+2\lambda }{4\lambda }},\qquad \mathcal{E}\sim
(t_{0}-t)^{\frac{1-2\lambda }{4\lambda }},\qquad \phi \sim (t_{0}-t)^{\frac{1+2\lambda }{4\lambda }}.  \label{32}
\end{equation}
During both accelerated expansions $H=A/t$ and $H=B/(t_{0}-t)$ energy
increases whereas the field $\phi $ decreases even though the Universe
expands. However, during the phase $H=A/t$ energy density decreases while
during the phase $H=B/(t_{0}-t)$ energy density increases. It is this crucial difference between the behavior of the elastic energy density during the two types of expansion that induces us to assign the former to the early-times while reserving the latter to late-times. We shall first give an argument to justify this choice and then describe a possible alternative in which the phantom behavior is viewed as a phantom-like inflation. First, here is the argument for the first choice. When elastic energy increases
at the same time that energy density decreases induces the subsequent
formation of radiation and matter in the form of defects and dislocations in
the continuum of space-time, i.e., the excess of elastic energy is
transformed into radiation and matter whose formation is favored both by
the energy density decrease and the decrease of $\phi $. This would be the
analogue in this model of the reheating process. The radiation and matter
are thus nothing but an elastic energy stored in the form of dislocations
and defects. When the density of these becomes non-negligible, the
subsequent evolution of the Universe is more properly described as we saw
above by the 'metric' gravity through the Friedmann-Lema\^{\i}tre equations.
When the density of matter then falls off considerably due to cosmic
expansion, the 'elastic' gravity turns on again, but this time starts out
with the value of $\phi $ acquired just before entering the phase $H=B/(t_{0}-t)$.
Since the subsequent decrease of $\phi $ is of the form $
(t_{0}-t)^{(1+2\lambda )/4\lambda }$, putting $t=0$ and equating that with
the value of $\phi $ acquired at the end of the matter-dominated era -- that
would be given by a model including dislocations properly -- would yield the
value of $t_{0}$.

Now, the latter form of expansion is usually ascribed to a phantom dark
energy, for which the Universe might actually go towards the Big Rip
singularity in the finite time $t_{0}$. The present model, however, suggests
a way out. Indeed, during such expansion we have a negative pressure that
pulls inwards, a deformation $\phi$ that decreases faster than the increase
of the scale factor, and also an energy density that increases faster than
the scale factor. Thus, when reaching the Planck scales quantum mechanical
effects would begin to dominate before the scale factor has any chance to
blow up. It is then not excluded that a quantum mechanical behavior of the
field $\phi$ may prevent the Universe from following its late-time dynamical
regime until it reaches the singularity.

The second possibility alluded to above is to assign the solution $H=B/(t_{0}-t)$, and hence the behavior (\ref{23}) of the scale factor, to a phantom inflation while reserving the power-law behavior for the late-time expansion. Indeed, phantom inflation has also gained considerable attention in the literature and the smallness of the exponent in $a\propto (t_{0}-t)^{-0,07}$ obtained above agrees quite well with the requirements expected from a phantom-like inflation (see e.g. \cite{31}). With this interpretation, however, it is the increase in the energy density during the phase $H=B/(t_{0}-t)$ that should be taken responsible for the formation of dislocations and ascribed to the reheating process. However, lacking a real model that would describe matter as space-time dislocations, we shall not pursue this issue further here.

\section{Concluding remarks}\label{Concl}
In this paper we studied the dynamics of the Universe in the
framework of the assumption that space-time is an elastic continuum whose
deformations obey generalized equilibrium equations of three-dimensional
elasticity and whose entropy is always extremal. We have seen that it is
possible to recover both the inflationary expansion of the Universe and a
late-time phantom-like behavior without appealing to inflaton and phantom
fields. We have also argued for the possibility for a phantom inflation to arise naturally within this model. These do not, however, exhaust all the possibilities one might expect to achieve within the present model, because there is yet a third alternative, which is to have a late-time de Sitter Universe and a cosmological constant. Indeed, when $\eta=0$ (i.e., when the Poisson's ratio is exactly equal to 0.2) the dynamical equation (\ref{19}) becomes $-\ddot{H}H+\frac{3}{2}\dot{H}^{2}+3\lambda\dot{H}{H^{2}}=0$. This equation has two positive solutions, the first is a constant Hubble parameter $H=H_{0}$ whereas the other is again $H=B/(t_{0}-t)$ with $B=1/6\lambda$. As we have remarked above, though, the latter supplies an insufficient phantom-like behavior for a late-time expansion. If, however, one interprets again this solution as a phantom inflation for the very early Universe then the other constant solution would naturally describe a late-time de Sitter Universe with an eternal exponential expansion $a\sim\exp(H_{0}t)$ during which the elastic energy density remains constant.

Another great merit of the scalar field-based inflationary scenario,
however, is its possibility to provide also the observed large scale structure of
the Universe by studying the primordial quantum fluctuations of the inflaton
field that get magnified to the presently observed ones through the rapid
expansion of the very early Universe \cite{5}. It is, in principle, also
possible to conduct a similar analysis within the present model by
considering the quantum fluctuations of the scalar $\phi$. But since the
latter is still viewed as a classical field in our approach, the study of
its eventual quantum mechanical origins and quantum fluctuations are beyond
the scope of the present work and will be deferred to future investigations.

Finally, we see that a vanishing or a decreasing scale factor $a(t)$ are
both forbidden in the framework of our present model since Eq.~(\ref{17})
would then make entropy either vanish or come out negative. This constitutes
a hint that the original Big Bang singularity may also be avoided within
this approach. But since that issue pertains to the realm of the Planck era,
a proper treatment of that question may again only come from a quantum
mechanical model for the field $\phi$.

\end{document}